\documentclass[preprint,aps,prb, amsmath,amssymb,superscriptaddress,showpacs]{revtex4-1}
\usepackage{graphicx}% Include figure files
\usepackage{dcolumn}% Align table columns on decimal point
\usepackage{bm}% bold math
\begin{document}

\preprint{}%APS/PRB draft}

\title{Multidimensional Coherent Spectroscopy of a Semiconductor Microcavity}

\author{Brian~L.~Wilmer}
\affiliation{Department of Physics and Astronomy, West Virginia
University, Morgantown, WV, 26501-6315, USA}

\author{Felix~Passmann}
\affiliation{Department of Physics and Astronomy, West Virginia
University, Morgantown, WV, 26501-6315, USA}
\affiliation{Experimentelle Physik 2, TU Dortmund, Dortmund,
Germany 44221}

\author{Michael~Gehl}
\affiliation{College of Optical Sciences, The University of
Arizona, Tucson, AZ 85721-0094, USA}

\author{Galina~Khitrova}
\affiliation{College of Optical Sciences, The University of Arizona, Tucson, AZ 85721-0094, USA}

\author{Alan~D.~Bristow}
\email{E-mail: alan.bristow@mail.wvu.edu} \affiliation{Department
of Physics and Astronomy, West Virginia University, Morgantown,
WV, 26501-6315, USA}

\begin{abstract}
Rephasing and non-rephasing two-dimensional coherent spectra map
the anti-crossing associated with normal-mode splitting in a
semiconductor microcavity. For a 12-meV detuning range near zero
detuning, it is observed that there are two diagonal features
related to the intra-action of exciton-polariton branches and two
off-diagonal features related to coherent interaction between the
polaritons. At negative detuning, the lineshape properties of the
diagonal intra-action features are distinguishable and can be
associated with cavity-like and exciton-like modes. A biexcitonic
companion feature is observed, shifted from the exciton feature by
the biexciton binding energy. Closer to zero detuning, all
features are enhanced and the diagonal intra-action features
become nearly equal in amplitude and linewidth. At positive
detuning the exciton- and cavity-like characteristics return to
the diagonal intra-action features. Off-diagonal interaction
features exhibit asymmetry in their amplitudes throughout the
detuning range. The amplitudes are strongly modulated (and invert)
at small positive detuning, as the lower polariton branch crosses
the bound biexciton energy determined from negative detuning
spectra.
\end{abstract}

\date{\today}
%\keywords{Microcavity, semiconductors, coherent spectroscopy, nonlinear optics}

\maketitle

Semiconductor microcavities supporting
exciton-polaritons\cite{1a,1b} are used in optoelectronic
application\cite{1,2,3} and provide a platform for exploring
exotic coherent physical phenomena.\cite{4,5,6,7,8,9,10,11,12} The
normal-mode coupling between the photonic cavity mode ($\gamma$)
and the exciton resonance (X) enhances both the linear and
nonlinear optical interactions.\cite{13a} Transient four-wave
mixing (FWM) confirm that strong exciton-cavity interactions
modify the temporal behavior of the coherent response and
many-body Coulomb correlations determine the exact
dynamics,\cite{13} which affects dissipation\cite{5,14,15} and
coherent control.\cite{16,17} Biexciton-polaritons also contribute
to the overall emission signal, even through the biexciton binding
energy is only slightly altered by the cavity.\cite{18,19}

Multidimensional coherent spectroscopy (MDCS) is based on and
supersedes FWM. At optical frequencies, MDCS has been utilized to
study various semiconductor
nanostructures.\cite{20,21,22,23,24,25,26} This technique retains
both time and frequency resolution, is able to unambiguously
distinguish a variety of quantum pathways (including those with
non-radiative steps),\cite{27} and can separate homogeneous and
inhomogeneous broadening.\cite{28} To date, two-quantum, and
higher-order,\cite{29} coherent spectra have shown that many-body
interactions dominate the signals, including a contribution from
bound biexcitons for excitation with the correct polarization
configuration. Two-dimensional coherent spectroscopy (2DCS) has
also examined coherence and control of excitonic qubits in
microcavity pillars.\cite{29a}

Despite the extensive fundamental and applied studies of
microcavity exciton-polaritons, the anti-crossing has not been
systematically mapped using MDCS. In this paper, normal-mode
splitting of a semiconductor microcavity and the associated
exciton-polariton branches are mapped using rephasing and
non-rephasing 2DCS. This study is performed over a range of energy
detuning ($\Delta = E_{\gamma}-E_{\rm X}$) near the anti-crossing,
where $E_{\gamma}$ and $E_{\rm X}$ correspond the cavity-mode and
exciton-mode energies respectively. The detuning-dependence of
spectral features, related to intra-action (diagonal features) of
and interaction (off-diagonal features) between polariton
branches, informs us about the coupling between the cavity and
excitonic or biexcitonic modes. For example, a contribution from a
bound biexciton is isolated at negative detuning, which is
convolved with the off-diagonal features at positive detuning and
strongly modulates their relative amplitudes.

\begin{figure}[b]
\centering{\includegraphics[width=8cm]{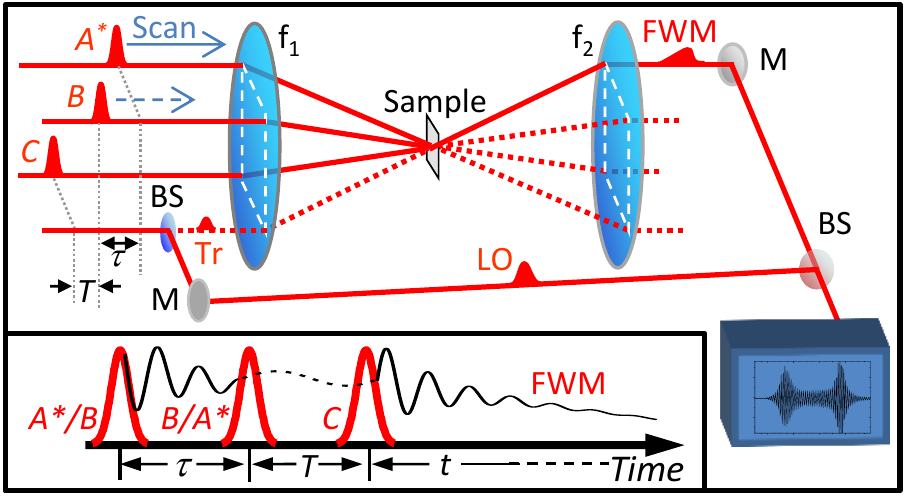}} \caption{(Color
online) Experimental setup for the multidimensional coherent
spectroscopy. (f = lens, M = mirror, BS = beam splitter, FWM =
four-wave-mixing signal, $\tau$ = period between pulses A and B,
$T$ = period between pulses B and C, and $t$ = period after pulse
C triggers the emission, Tr = tracer beam, traces the
phase-matched direction, and LO = local oscillator for spectral
interferometry). The inset shows the excitation pulse sequence.}
%\label{fig:fig1}
\end{figure}

The experimental setup is described fully elsewhere.\cite{30} In
brief, the laser source is a mode-locked Ti:Sapphire oscillator
that produces 100-fs pulses. A MONSTR is used to create and phase
control four identical pulses arranged on the corners of a box. As
shown in Fig.~1, three pulses impinge the microcavity sample,
which resides in an optical cryostation at the focus and crossing
point of the beams. A tracer (Tr) beam is used for alignment and
blocked for the FWM and 2DCS measurements. All measurements are
performed in the third-order nonlinear optical regime, excited
with average powers of $0.2 - 0.7$~mW per beam. In a third-order
perturbation excitation scheme, the signal is generated from
interaction by all three excitation pulses. The excitation
sequence is shown in the inset of Fig.~1: the first pulse creates
a coherent superposition between the ground and excited states,
the second pulse then creates a population in either the ground or
excited state, and the third pulse converts the population into a
radiating polarization. This polarization is emitted as a
transient FWM signal, which is collected in transmission mode and
directed to a spectrometer and CCD camera.

The microcavity sample (denoted NMC73) was grown by molecular beam
epitaxy on a GaAs substrate.\cite{1a} The mirrors consist of
GaAs/AlAs (14.5 and 12 bilayer) distributed Bragg reflectors
separated by a wedged $\lambda$ GaAs cavity, with a cavity mode
close to 830~nm. In the center of the cavity, at the antinode of
its electric field, is a single 8-nm In$_{0.04}$Ga$_{0.96}$As
quantum well. The reflection properties at low temperature (not
shown) exhibit the typical normal-mode splitting expected for such
a structure. Translating the sample detunes the cavity mode with
respect to the bare exciton energy.

\begin{figure}[b]
\centering{\includegraphics[width=8cm]{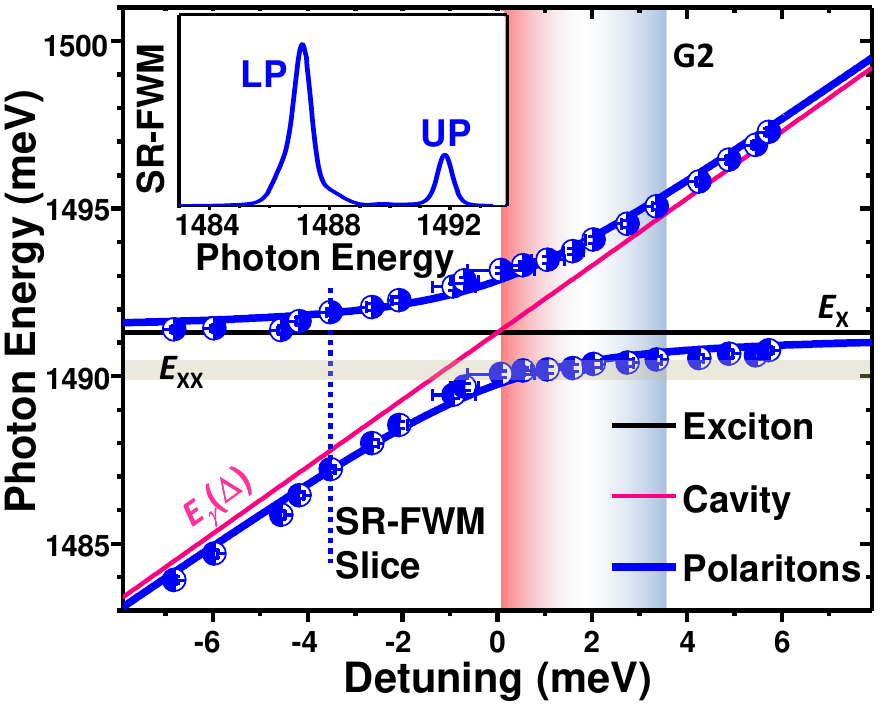}} \caption{(Color
online) Detuning dependence of the features observed in spectrally
resolved four-wave mixing (FWM). A typical FWM spectrum is shown
in the inset, from which the center energies of the lower (LP) and
upper (UP) polaritons are extracted for the body of the figure.
Solid lines model the vacuum Rabi splitting, the bare exciton
energy and the detuning for each spectrum. Also shown are the
expected energy of the bound biexciton, $E_{\rm XX}$ and the
detuning range of the G2 profile discussed in Fig.~4.}
%\label{fig:fig2}
\end{figure}

Spectrally resolved FWM is acquired as a precursor to performing
2DCS with $\tau = 0$~fs and $T = 100$~fs. The inset of Fig.~2
shows a typical spectrum, revealing resonances associated with the
lower (LP) and upper (UP) polariton branches. Figure.~2 shows the
spectral positions extracted from spectra measured at different
positions on the sample. The detuning, $\Delta$, is determined for
each spectrum by modeled the data using
\begin{equation}
 E_{UP/LP}(\Delta) = 1/2[2E_{\rm X} + \Delta \pm (\Delta^{2} + E_{\rm VRS}^{2})^{1/2}],
\end{equation}
where $E_{\rm VRS}$ is the vacuum Rabi splitting, a measure of the
coupling between the optical field and the excitons. It is found
that $E_{\rm VRS} = 3.1\pm0.1$~meV and $E_{\rm X} =
1491.3\pm0.03$~meV, which both agree well with results on similar
structures.\cite{1a} The inset spectrum is then indicated as a
dashed vertical line at $\Delta \simeq -3.75$~meV. Microcavities
are sensitive to the angle of incidence, since the dispersion
varies with the in-plane wavevector. Here the external angle of
incidence for each beam is approximately 7$^{\rm o}$ and the axis
of the box of pulses is at normal incidence. Each beam couples to
slightly different polariton states in momentum space, which may
lift the measured energy and lower the overall signal strength.
However, close to zero detuning, the FWM emission requires
attenuation of at least ND2.5, indicating minimal momentum-related
signal reduction and excitation within the parabolic region of the
in-plane momentum.\cite{31}

For 2DCS measurements, the transient FWM is collected in the
spectrometer along with a phase-stabilized local oscillator (LO)
pulse, such that complex spectra can be recorded by spectral
interferometry. Spectra are acquired for a range of time delays,
$\tau$, scanned in phase-stabilized increments. A numerical
Fourier transform is performed to convert $\tau$ to
$\omega_{\tau}$. If the conjugate pulse $A^{*}$ is scanned
backward in time (toward the sample), the time-ordering results in
the phase-matching condition $k_{s}=-k_{A^{*}}+k_{B}+k_{C}$.
Transient FWM exhibits a photon echo for inhomogeneously broadened
systems. This method records a rephasing 2D spectrum,
$S_{I}(-\omega_{\tau},T,\omega_{t})$, wherein the diagonal
($\hbar\omega_{\tau}=\hbar\omega_{t}$) of the plot is towards the
lower right corner, due to the numerical choice of the emission
photon energy. Rephasing spectra allow for the separation of
homogeneous and inhomogeneous linewidths. Alternatively, if pulse
$B$ is scanned instead of pulse $A^{*}$, then the phase matching
becomes $k_{s}=k_{B}-k_{A^{*}}+k_{C}$ and the spectrum is
non-rephasing, $S_{II}(\omega_{\tau},T,\omega_{t})$.

Figure~3 shows the absolute field amplitude of the rephasing
(bottom row) and non-rephasing (top row) 2DCS results for a range
of detuning values from (a) $\Delta = -5$~meV to (e) $+4.5$~meV.
Excitation is performed with a mixing time $T = 100$~fs and
collinear polarization (XXXX), where the notation corresponds to
the polarization state of the three pump pulses and the emission.
Each panel is normalized to the strongest peak for presentation.
At each value of $\Delta$ the laser spectrum is overlaid with the
non-rephasing spectrum, illustrating that the two resonances are
excited equally in each case. This is important for careful
comparison of the relative amplitudes of each feature as a
function of detuning. Due to the transmission geometry, strong
absorption of the tracer beam prevents experimental determination
of the global phase using all-optical methods or via
spectrally-resolved transient absorption.\cite{32} Hence, only
amplitude spectra are shown.

In Fig.~3(a) the $\gamma$-like mode is the low-energy feature,
denoted A, and is broader than the higher-energy X-like mode,
denoted B. From analysis of the lineshapes the cross-diagonal
width is dominated by the homogeneous linewidths\cite{32a} of
intra-action features A and B, yielding values of $\gamma_{\rm LP}
= 0.35$~meV and $\gamma_{\rm UP} = 0.14$~meV respectively. In
comparison, the diagonal linewidths, which can be dominated by
inhomogeneous broadening, are only slightly wider in each case:
$\sigma_{\rm LP} = 0.38$~meV and $\sigma_{\rm UP} = 0.17$~meV. In
addition to the diagonal intra-action features, two off-diagonal
interaction features C and D are observed, which are due to
coherent coupling between the A and B features.\cite{20}

%\onecolumngrid
\begin{figure}
\centering{\includegraphics[width=16cm]{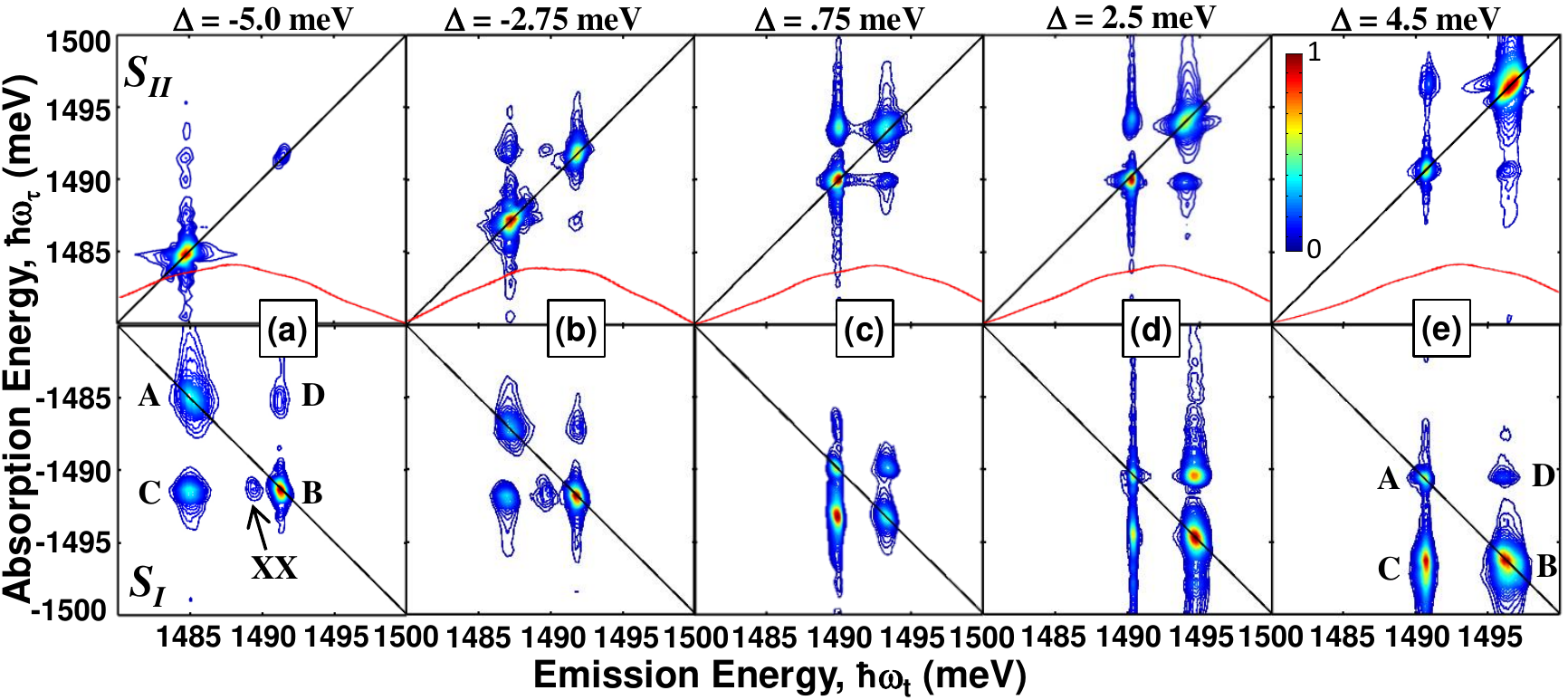}} \caption{(Color
online) Rephasing, $S_{I}$ (bottom row), and non-rephasing,
$S_{II}$ (top row), two-dimensional coherent spectra for the
labelled range of detuning. Also shown is the laser excitation
spectrum for each excitation position. Diagonal intra-action
features are A and B, and off-diagonal interaction features are C
and D. Negative detuning also shows a biexciton feature, XX.}
%\label{fig:fig3}
\end{figure}
%\twocolumngrid

Increasing $\Delta$ shifts all spectral features toward higher
energy. The separation between the A and B modes (projected onto
the emission axis) is $\sim 7$~meV in (a), decreases as $\Delta$
tends to zero, becoming $~3.4$~meV in (c), and increases again to
$~5.6$~meV in (e). Close to zero detuning ($\Delta=0.75$~meV), the
homogeneous linewidths of A and B in the $S_{I}$ spectra are
$\gamma_{\rm LP} = 0.12$~meV and $\gamma_{\rm UP} = 0.18$~meV
respectively. The properties are nearly identical and
$\gamma$-like and X-like characteristics are no longer
distinguishable, since it is expected that the observed linewidths
should become identical.\cite{32b} For larger positive detuning
the homogeneous linewidths of A and B are $\gamma_{\rm LP} =
0.14$~meV and $\gamma_{\rm UP} = 0.24$~meV respectively. The
$\gamma$-like and X-like characteristics are once again
distinguishable, but are not quite the same as for negative
detuning. As expected from analysis of the Hopfield
coefficients,\cite{18} the modes switch and the $\gamma$-like mode
is now the upper polariton, B. The mode switch is consistent with
results where the mirror reflectivity results in a narrower cavity
than bare exciton linewidth.\cite{32b}

Collinear polarization in 2DCS allows for excitation to the
biexciton (XX) states.\cite{33} For bound biexcitons, binding
energy acts to shift the XX feature laterally from the X-like mode
in the emission energy, $\hbar\omega_{t}$. This feature is only
observed for negative detuning, from which the XX binding energy
is determined to be $\sim1.88$~meV. The biexciton follows the
expected excitation-density dependence and is suppressed for
co-circular polarization (data not shown). Its cross-diagonal
linewidth is almost identical to that for the X-like mode, which
is expected because the quantum pathway that creates the XX
feature is a two-step excitation via the exciton. Hence, the
linewidth projected on the absorption energy,
$-\hbar\omega_{\tau}$, should be identical to that for the
exciton. The linewidth projection onto $\hbar\omega_{t}$ may be a
little wider (tilting the feature away from the diagonal),
depending on the degree of correlation of the exciton and
biexciton states.\cite{34,35} In this case, the exciton and its
biexciton are parallel and are highly correlated.

Non-rephasing, $S_{II}$, are presented for comparison, showing
very similar results to $S_{I}$ spectra across the entire detuning
range. $S_{II}$ spectra typically have slightly weaker
off-diagonal features, as is observed here. Otherwise, the two
diagonal intra-action, two off-diagonal interaction and biexciton
features are all observed as discussed above.

Figure~4 shows the integrated amplitude versus detuning for the
four main polaritonic features A through D. In each case, a small
area around each feature is integrated. Panel (a) shows the total
integrated amplitude for a 12-meV detuning range and is normalized
to the highest emission strength at $\Delta = 0$~meV. Significant
enhancement of the exciton-polariton transition are observed due
to normal-mode coupling,\cite{36} as the integrated amplitude
increases and peak close to zero detuning. The peakshape of
extracted total amplitude is asymmetric in detuning\cite{37} and
can be fit with two gaussian lineshapes: G1 is centered at zero
detuning and G2 is centered at 1.88~meV (G2), corresponding to the
biexciton binding energy determined from negatively detuned
spectra. Fitted full widths at half maximum of G1 and G2 are $\sim
3.39$~meV and $\sim 3.65$~meV respectively. The sum of G1 and G2
well represent the $\Delta$-dependence peakshape.

\begin{figure}[t]
\centering{\includegraphics[width=8cm]{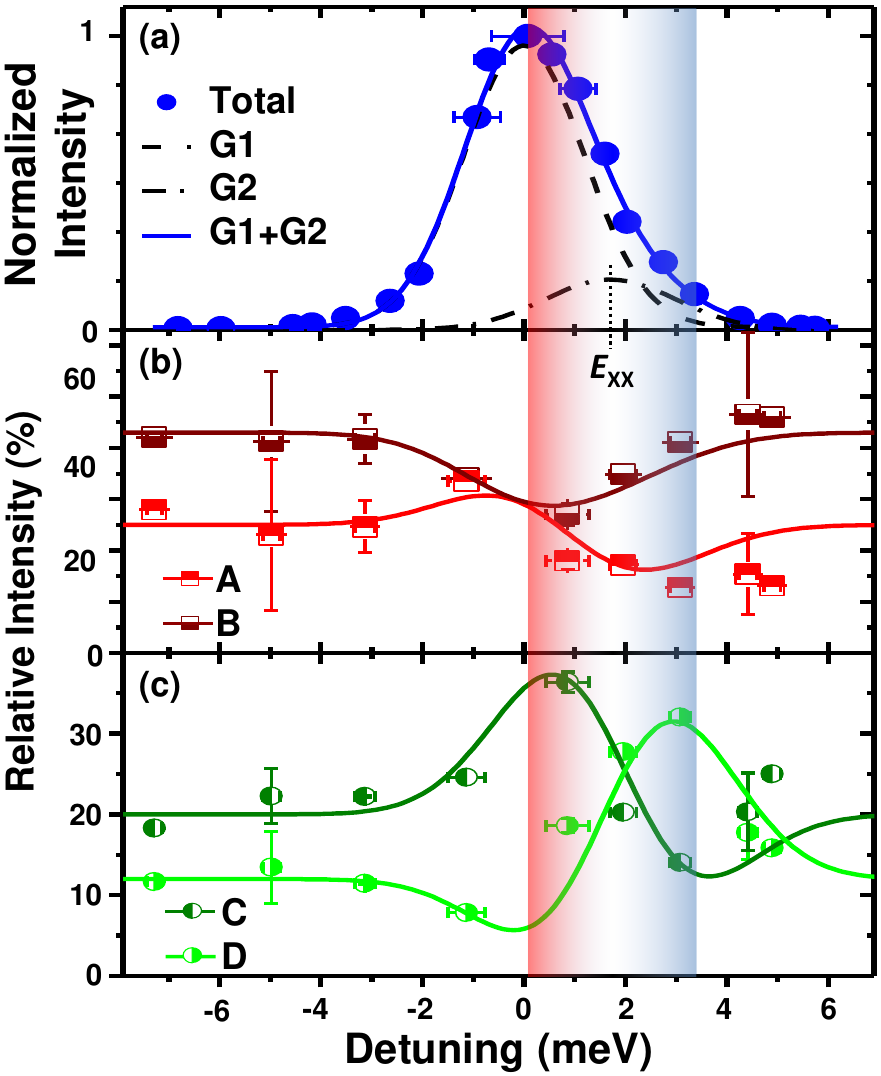}} \caption{(Color
online) Detuning dependence of the features in the rephasing 2DCS.
(a) Shows the normalized amplitude of the entire spectrum, with
best fits based on two gaussian profiles G1 and G2. (b) Shows the
relative amplitude of the diagonal features A and B. (c) Shows the
relative amplitude of the off-diagonal features C and D. The solid
lines in (b) and (c) model the affect of the two resonances
parameterized by the G1 and G2 gaussian profile in (a).}
%\label{fig:fig4}
\end{figure}

Figure~4 (b) and (c) show $\Delta$-dependence of the relative
amplitude for the individual diagonal intra-action and
off-diagonal interaction spectral features. From each spectrum the
individual features' integrated amplitudes are extracted and
normalized to the total integrated amplitude. At zero detuning the
relative amplitudes of the diagonal intra-action features are
identical, showing that normal-mode coupling leads to enhancement
and equalization of the UP and LP branches. Away from zero
detuning the higher-energy diagonal feature, B, is always stronger
and the overall amplitude of the diagonal features drops at small
positive detuning. In contrast, the relative integrated amplitude
of the off-diagonal features is smaller at negative detuning and
oscillates for positive detuning. The oscillation position is
consistent with the position of the small G2 peak from panel (a).
In this range, the amplitude of peak C increases rapidly,
corresponding to the increasing slope of G2, and decreases rapidly
to become smaller than D, corresponding to the decreasing slope of
G2. Both the width of G2 and the range of $\Delta$ where the
oscillations occurs agree well with the width of the LP branch
passing through the bound biexciton; see Fig.~2.

The solid lines in Fig.~4(b) and (c) model the relative amplitudes
of the diagonal and off-diagonal features based on the following
parameters. First, amplitude offsets for each feature are selected
from negative detuned spectra, yielding 25~$\%$, 43~$\%$, 10~$\%$
and 12~$\%$, for A, B, C and D respectively. Second, near $\Delta
= 0$~meV, the strength of the two diagonal features equalizes,
bringing the two modes close in amplitude in the range of the G1
profile. This $\sim9$\% deviation is symmetric for each feature.
Third, in the region of the G2 profile $\sim11$\% of the spectral
weight is transferred from both diagonal features to the
off-diagonal features, most likely due to additional many-particle
interaction terms in the quantum pathways to the biexciton
manifold. Fourth, the off-diagonal features experience a $\pm
d({\rm G2})/d\Delta$ modulation in their spectral weight by
approximately $\pm14$\%, which is most likely due to a attraction
or repulsion of LP branch as it passes through XX. The latter two
effects are complementary evidence of the recently observed
Feshbach resonance.\cite{12} Feshbach resonances occur when the
energy of two free, yet interacting, polaritons is in resonance
with the bound molecular excitonic state. Off-diagonal features
are the interaction between the LP and UP branches, so that even
though the LP branch alone overlaps with the XX, both off-diagonal
amplitudes are modulated and invert. This result arises from
coherent coupling by quantum interference of the polaritons
through the shared ground state or by a Raman-like coherence
between the excited polaritons.\cite{27} 2DCS sensitivity is
revealed, because no splitting is observed associated with the
Feshbach resonance in the linear spectra, yet the influence of
this LP-XX crossing is clear.

In summary, this study has mapped the detuning dependence of the
cavity mode through the exciton and biexciton modes of a single
quantum well and isolated the coherent response using
two-dimensional coherent spectroscopy. Enhancement of the
four-wave mixing emission was observed near zero detuning, along
with anti-crossing of the upper and lower polariton branches.
Homogeneous and inhomogeneous linewidths are found to be
consistent with those for a wider cavity than bare exciton mode
Amplitudes of the spectral features are highly sensitive to the
interaction between exciton, biexciton and cavity modes, revealing
strong modification as bands intersect. This work paves the way
for determining contributions through polarization- and
excitation-dependent studies using 2DCS and begs full microscopic
theoretical treatment to reproduce spectral features. Moreover,
these methods can be used to disentangle the coherent and
transient phenomena that parallel processes identified in
ultracold atomic physics, such as condensation and superfluidity.

$~$

The authors wish to acknowledge Steven Cundiff for useful
discussions. The work at WVU was supported by the National Science
Foundation (CBET-1233795) and the WV Higher Education Policy
Commission (HEPC.dsr.12.29). Work at Arizona was supported by
AFOSR (FA9550-13-1-0003), NSF-AMOP and NSF REC-CIAN.

\end{document}